\documentclass[preprint]{aastex61}
\usepackage{amsmath}
\usepackage{graphicx}
\usepackage{color}

\shorttitle{Testing isotropy of log \textit{N} - log \textit{S} slope} 
\shortauthors{Ghosh \& Jain}
\submitjournal{ApJ}

\begin{document}
\title{Testing the isotropy of the log \textit{N} - log \textit{S} slope for the NVSS radio catalogue}
\correspondingauthor{Shamik Ghosh}
\email{shamik@iitk.ac.in}

\author{Shamik Ghosh}
\affiliation{Department of Physics, \\Indian Institute of Technology, Kanpur, \\U.P. 208016, India}

\author{Pankaj Jain}
\affiliation{Department of Physics, \\Indian Institute of Technology, Kanpur, \\U.P. 208016, India}

\begin{abstract}
The cumulative number count $N$, of sources above a threshold is known to 
approximately follow a power law behaviour $N\propto S^{-x}$. We study the variation of  spectral index $x$ across the sky in order to look for possible 
signals of violation of isotropy. 
We develop a rigorous algorithm of likelihood maximisation to accurately fit for the spectral index. We  
divide the sky into upper and lower 
hemispheres for a particular choice of $z$-axis and determine 
the difference $\Delta x$ between the best fit values of the spectral indices 
between the two hemispheres. The maximum value of this difference obtained
by varying over
the $z$-axis provides us with a measure of departure from isotropy. We
find that the data support isotropy of the spectral index. 
The maximum difference is found to be
 1.3\% of the full sky best fit value of $x$. 
The deviation
is found to be significant only at 2$\sigma$ level which indicates a  
weak departure from isotropy. We also perform a dipole fit to the spectral
index as a function of the angular coordinates. The result is found to be
consistent with isotropy. 
\end{abstract}


\section{Introduction}
The cumulative number count $N$, of sources above a threshold flux density $S$ per unit solid angle is typically assumed to have a power law behaviour. The slope $x$ of the $\log N - \log S$ curve has played an important role in 
astronomy and in early cosmology. The variation of the $\log N$  with $\log S$ can be related
to the underlying cosmological model \citep{Davidson:1962,Longair:1966,Petrosian:1969}. Observationally a power law behaviour is found to provide a 
reasonably good fit to data, although deviations from a pure power law behaviour
have been observed \citep{Condon:1998,Kothari:2013,Tiwari:2013}. It is generally assumed
that slope is independent of the direction and position of observation. 
This follows from the cosmological principle. 
However there currently exist
several signals which suggest potential  
 violations of statistical isotropy. These include 
dipole anisotropy in radio polarization angles \citep{Jain:1998}, radio source counts
and brightness \citep{Blake:2002,Singal:2011,Gibelyou:2012,Rubart:2013,Kothari:2013}, radio polarized source counts and brightness \citep{Tiwari:2013}, alignment
of Cosmic Microwave Radiation (CMB) quadrupole and octopole \citep{deOliveiraCosta:2003}, CMB hemispherical anisotropy \citep{Eriksen:2004}, alignment of
quasar polarizations \citep{Hutsemekers:1998,Jain:2003sg} etc.  
Although not confirmed, these signal are very interesting and
are being extensively studied in the literature.
Hence it is important to test whether the slope $x$
 is consistent with the principle of isotropy.  
In this work we study the direction dependence of this parameter.

The number counts as well as brightness of radio sources acquire 
a dipole distribution due to several contributions. These include
our local motion relative to the cosmic frame of rest as well as the 
dipole arising due to local clustering of sources. The
 slope $x$ provides us with an
independent and potentially more robust measure to test the cosmological 
principle. 
In particular it is unaffected by the local motion. Furthermore it is a
probe of the flux distribution of sources rather than the number 
density and hence many factors which affect the dipole in number counts
may not contribute to $x$ or may contribute differently.
These factors include the local cluster sources as well as generalized
cosmological models which may assume anisotropic and/or inhomogeneous
Universe. 

In \textsection\ref{sec:theory} we discuss the cumulative and differential number count distribution functions and the definition of the likelihood functions used in our data analysis. We discuss the NVSS catalogue and its systematics and their removal in \textsection\ref{sec:data}. In \textsection\ref{sec:analysis} we discuss our data analysis pipeline and data simulation processes. The results are presented in \textsection\ref{sec:results}. 
\section{Theory}
\label{sec:theory}
The cumulative number count of radio sources per unit solid angle, $N_{>S}$, above a flux density of S can be expressed as \citep{Ellis:1984}:
\begin{equation}
N_{>S}\propto S^{-x}.
\label{eq:cumulative}
\end{equation}
The slope of the $\log N - \log S$ relation is equal to the spectral index $x$.
 It is easy to see that the differential number count per unit flux density per unit solid angle, $n(\theta,\phi,S)$, along a direction of observation $(\theta, \phi)$, assuming isotropy, is given by a power law form 
\begin{equation}
    n(\theta,\phi,S) = \frac{d^2N}{d\Omega dS} = k S^{-1-x}.
    \label{eq:distribution}
\end{equation}
Here $k$ is a normalisation constant dependent on the total number of sources, the flux range of the data and the solid angle. We are interested in testing the variation in the spectral index $x$ along different directions of the sky.

We use likelihood analysis in order to
fit the value of the spectral index. If the differential number count is integrated over solid angle and flux intensity then we should get the total number count. Let us integrate equation (\ref{eq:distribution}) over the full sky and a flux density range of $S_{\text{min}}$ and $S_{\text{max}}$. We get,
\begin{equation}
   \int_{\Omega_r}\int_{S_{\text{min}}}^{S_{\text{max}}} n(\Omega,S)dSd\Omega=\frac{\Omega_r k}{x} \left[S_{\text{min}}^{-x}-S_{\text{max}}^{-x}\right]=N_r,
\end{equation}
where $N_r$ is the number of sources in the flux range, $S_{\text{min}}$ to $S_{\text{max}}$, in a region of the sky which subtends a solid angle $\Omega_r$. This relation fixes  the normalization constant $k$. 

We define a probability $P(S|x)$ as the probability that any random radio source observed in the sky has a flux density of $S$. It is given by
\begin{equation}
    P(S|x)=\frac{1}{N_r}\int_{\Omega_r}n(\Omega,S)d\Omega = \frac{x}{\left[S_{\text{min}}^{-x}-S_{\text{max}}^{-x}\right]}S^{-1-x}=f(S).
    \label{eq:probability}
\end{equation}

We define our likelihood function as $\mathcal{L}=\prod_i^{N_r} P(S_i|x)$. Using equation (\ref{eq:probability}) we can write the log-likelihood as
\begin{align}
    \ln \mathcal{L}(x) &=\ln \left(\prod_i^{N_r} \frac{x}{\left[S_{\text{min}}^{-x}-S_{\text{max}}^{-x}\right]}S_i^{-1-x}\right) \nonumber \\
     &=N_r\ln \frac{x}{\left[S_{\text{min}}^{-x}-S_{\text{max}}^{-x}\right]}+(-1-x)\sum_i^{N_r}\ln S_i  
     \label{eq:likelihood}  
\end{align}
We maximize the log-likelihood given by equation (\ref{eq:likelihood}) to obtain the best fit value of spectral index $x$. We note here that this process does not incorporate the experimental errors in the measurement of the flux density $\sigma_i$. 

To account for the measurement error in flux density we write the observed flux density $F_i$ as
\begin{equation}
    F_i = S_i + e_i,
    \label{eq:delta_cond}
\end{equation}
where $S_i$ is the possible theoretical value drawn from distribution equation (\ref{eq:probability}) and $e_i$ is drawn from a standard normal distribution with variance given by $\sigma_i^2$, i.e.,
\begin{equation}
    P(e_i|\sigma_i) = \frac{1}{\sigma_i \sqrt{2 \pi}} \exp \left(-\frac{e_i^2}{2 \sigma_i^2}\right)=g(F_i-S_i).
    \label{eq:error_dist}
\end{equation}
We can then define the probability of an observed value of flux density $F_i$ as
\begin{align}
    P(F_i|x,\sigma_i) &= \int P(S_i|x) P(e_i|\sigma_i) dS_i = \int f(S_i) g(F_i-S_i) dS_i \nonumber\\
    &=\int\frac{x}{\left[S_{\text{min}}^{-x}-S_{\text{max}}^{-x}\right]}S_i^{-1-x}\frac{1}{\sigma_i \sqrt{2 \pi}} \exp \left(-\frac{(F_i-S_i)^2}{2\sigma_i^2}\right)dS_i
    \label{eq:prob_w_error}
\end{align}
using equations (\ref{eq:probability}), (\ref{eq:error_dist}) and (\ref{eq:delta_cond}). 

\begin{figure}[t]
    \centering
    \includegraphics[width=1.0\textwidth]{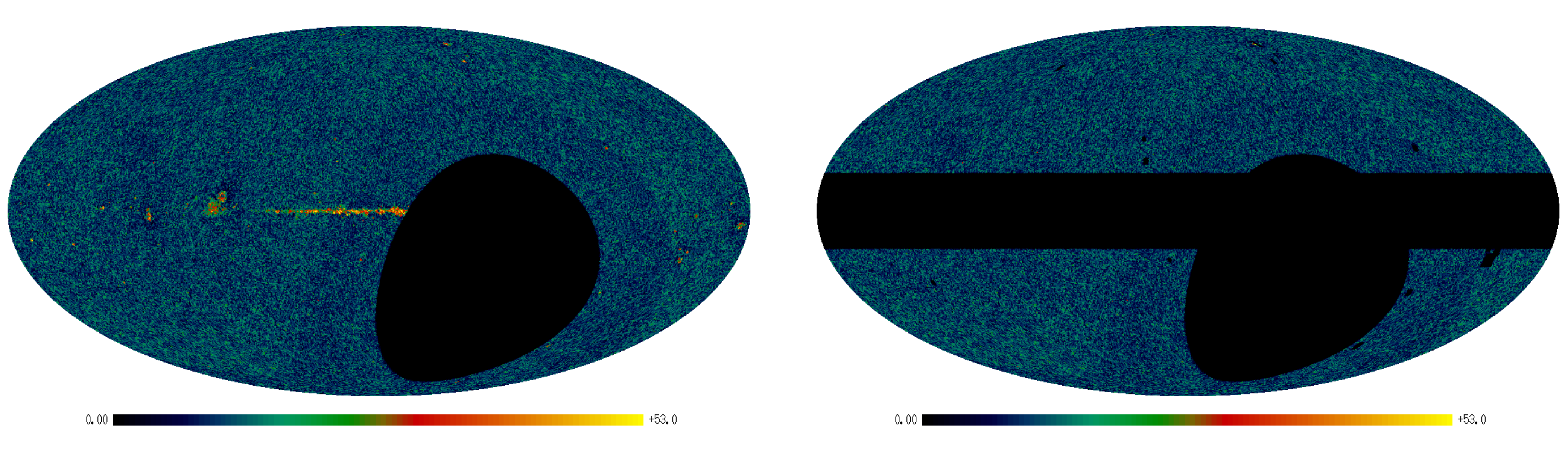}
    \caption{Left: Full NVSS catalogue with 82\% sky coverage. Right: NVSS catalogue with galactic plane and 22 extended bright sites masked with 56\% sky coverage. Both the maps are plotted in galactic coordinates and with a minimum of 0 and maximum of 53 (number of sources per pixel) in the colour scale. Note that both maps show a band of low number density below declination of $-10^\circ$.}
    \label{Fig:NVSS}
\end{figure}

The log-likelihood then is given by $\ln \mathcal{L} = \ln \left(\prod_i P(F_i|x,\sigma_i)\right)$. Using equation (\ref{eq:prob_w_error}) the form of the total log-likelihood is given by:
\begin{align}
    \ln \mathcal{L}(x) &= \sum_i^{N_r} \ln \left( \frac{x}{\left[S_{\text{min}}^{-x}-S_{\text{max}}^{-x}\right]}\frac{1}{\sigma_i \sqrt{2 \pi}}\int S_i^{-1-x} \exp \left(-\frac{(F_i-S_i)^2}{2\sigma_i^2}\right)dS_i \right) \nonumber \\
    &=N_r\ln\left(\frac{x}{\sqrt{2\pi}\left[S_{\text{min}}^{-x}-S_{\text{max}}^{-x}\right]}\right)+\sum_i^{N_r}\left[\ln \left\{ \int S_i^{-1-x} e^{-\frac{(F_i-S_i)^2}{2\sigma_i^2}}dS_i\right\}-\ln \sigma_i\right]
    \label{eq:L_w_error}
\end{align}
We use log-likelihood functions defined in equation (\ref{eq:likelihood}) and equation  (\ref{eq:L_w_error}) to fit for the spectral index $x$.

\section{Data}
\label{sec:data}
The NRAO VLA Sky Survey (NVSS) \citep{Condon:1998} is a continuum survey at 1.4 GHz covering the northern sky above a declination of $-40^\circ$ (J2000) with 82\% sky coverage. The NVSS catalogue contains 1773484 radio sources. 
We remove the galactic plane by removing all sources with galactic longitude $|b| < 15^\circ$. The galactic plane masking removes an additional 26\% of the sky. Hence we have only 56\% of the total sky area for analysis. The full catalogue with only the galactic portion masked is termed catalogue A for this work.

\subsection{Removing local clustering}
\citet{Blake:2002} had identified 22 sites of bright and extended radio galaxies. We follow their procedure of masking these 22 sites. This removes 23386 sources from the full NVSS catalogue. 
The total area removed is relatively small. 
As noted by \citet{Blake:2002}, the NVSS has poor resolution, so two or more galaxies situated near each other are rarely resolved into the individual galaxies. This is important as this would also increase the flux density for the unresolved object. Having many such unresolved clustered objects can cause deviation in the measured value of the spectral index.
This masked catalogue of 1750098 sources is further masked along the galactic plane, as discussed earlier, and the resulting data set is termed catalogue B. 

The expectation of isotropy is valid only when looking at cosmological length scales. The NVSS catalogue contains galaxies that are in our local neighbourhood and on these short distance scales local clustering would cause departure from isotropy. \citet{Blake:2002} have shown that the local clustering effects originate from redshifts of $z<0.03$. The Infrared Astronomical Satellite Point Source Redshift (IRAS PSCz) catalogue \citep{Saunders:2000} and the 3rd Reference Catalogue of bright galaxies (RC3) \citep{Vaucouleurs:1991,Corwin:1994} give a good sample for galaxies in the local supercluster. We prepare another catalogue which removes the NVSS sources which are within 30 arcsec separation from sources listed in IRAS PSCz catalogue and the RC3. This cut removes 13597 sources. We additionally mask the 22 bright and extended radio sources as discussed above. This keeps 1736583 sources. After masking the galactic plane as previously described we prepare catalogue C.

\begin{figure}[t]
    \centering
    \includegraphics[width=1.0\textwidth]{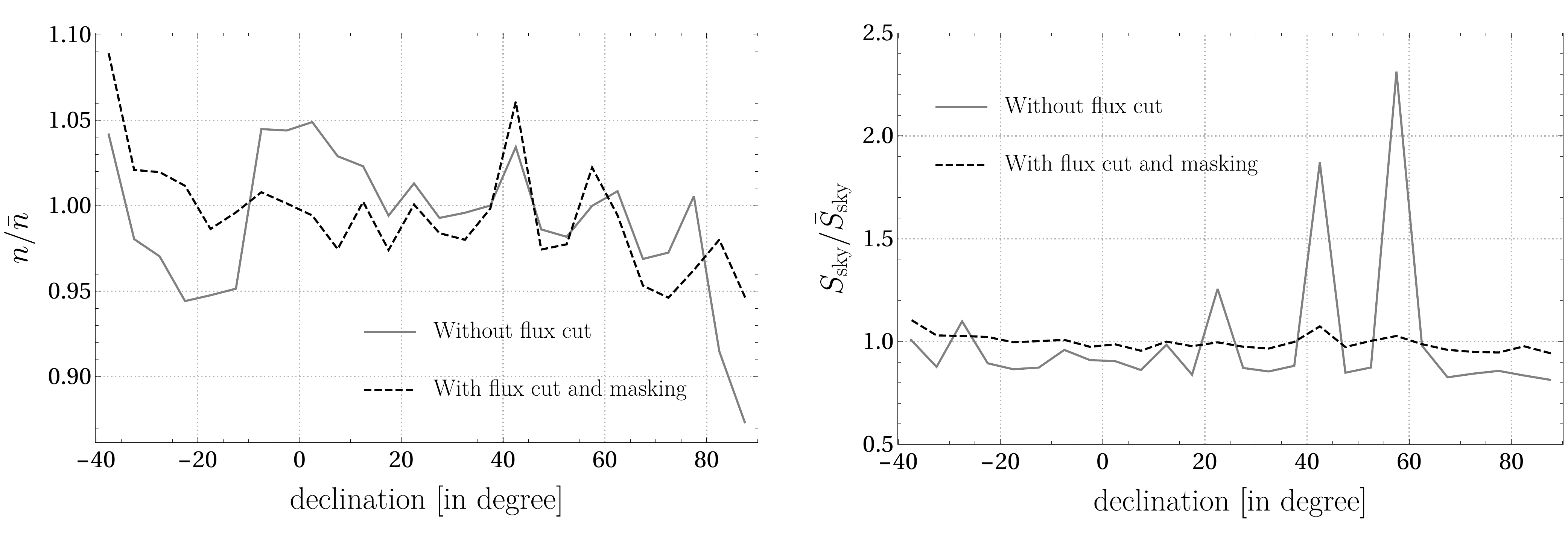}
    \caption{Left: The ratio of the local number count density to the whole sky average plotted with declination. Note the dip in the fraction below $-10^\circ$ and above $78^\circ$ where DnC configurations have been used. Right: The ratio of the local sky brightness to the whole sky average brightness plotted with declination. The spikes in the unmasked plot is due to the very bright sources densely packed around the galactic plane. In both plots bin sizes of $5^\circ$ has been used. The solid line shows the plot for the case without any flux cuts and the dashed line is for 20 mJy lower flux cut and a upper flux cut of 1000 mJy with the galactic plane masked.}
    \label{Fig:systematics}
\end{figure}

\subsection{Systematics}
The NVSS catalogue is well known to suffer from systematics. For the survey \citep{Condon:1998,Chen:2015} between the declinations of $-10^\circ$ and $78^\circ$ the D configuration of the VLA was used while for the remaining regions of the sky the DnC configuration was used. The D configuration apparently has more brightness sensitivity over the hybrid DnC configuration and hence the portion of the sky sampled with the D configuration has a higher source density. The number density of sources show clear variation with declination as can be seen in
 Figure \ref{Fig:NVSS}. 

To study the effects of configuration systematics we divide the sky in strips of constant declination with a width of $5^\circ$. We look at the ratio of the number density $(n)$ in each strip to the average number density $(\bar n)$ over the whole sky. We also define a quantity called sky brightness $(S_{\text{sky}})$, which is the sum of the flux densities of all the sources in a given portion of the sky divided by the area of the patch. We look at the ratio of the local sky brightness of a strip to the whole sky average brightness $(\bar{S}_{\text{sky}})$. The variations of these two ratios with declination is shown in Figure \ref{Fig:systematics}. From the plot of number density ratios it is very clear that below the declination of $-10^\circ$ and above the declination of $78^\circ$ there is clear dip in the number density of sources due to the DnC configuration used in surveying as discussed above. It has been shown by Blake and Wall that the declination dependence can be greatly reduced by placing a cut on the minimum flux of the sources to be considered. They showed that over a flux density of 15 mJy the effects due to 
limited brightness sensitivity do not remain an issue and the number density evens out. For our analysis we choose 20 mJy as the minimum flux density. We also place a cut on the maximum flux density of 1000 mJy to remove sources with large brightness.

 \begin{figure}[t]
     \centering
     \includegraphics[width=1.0\textwidth]{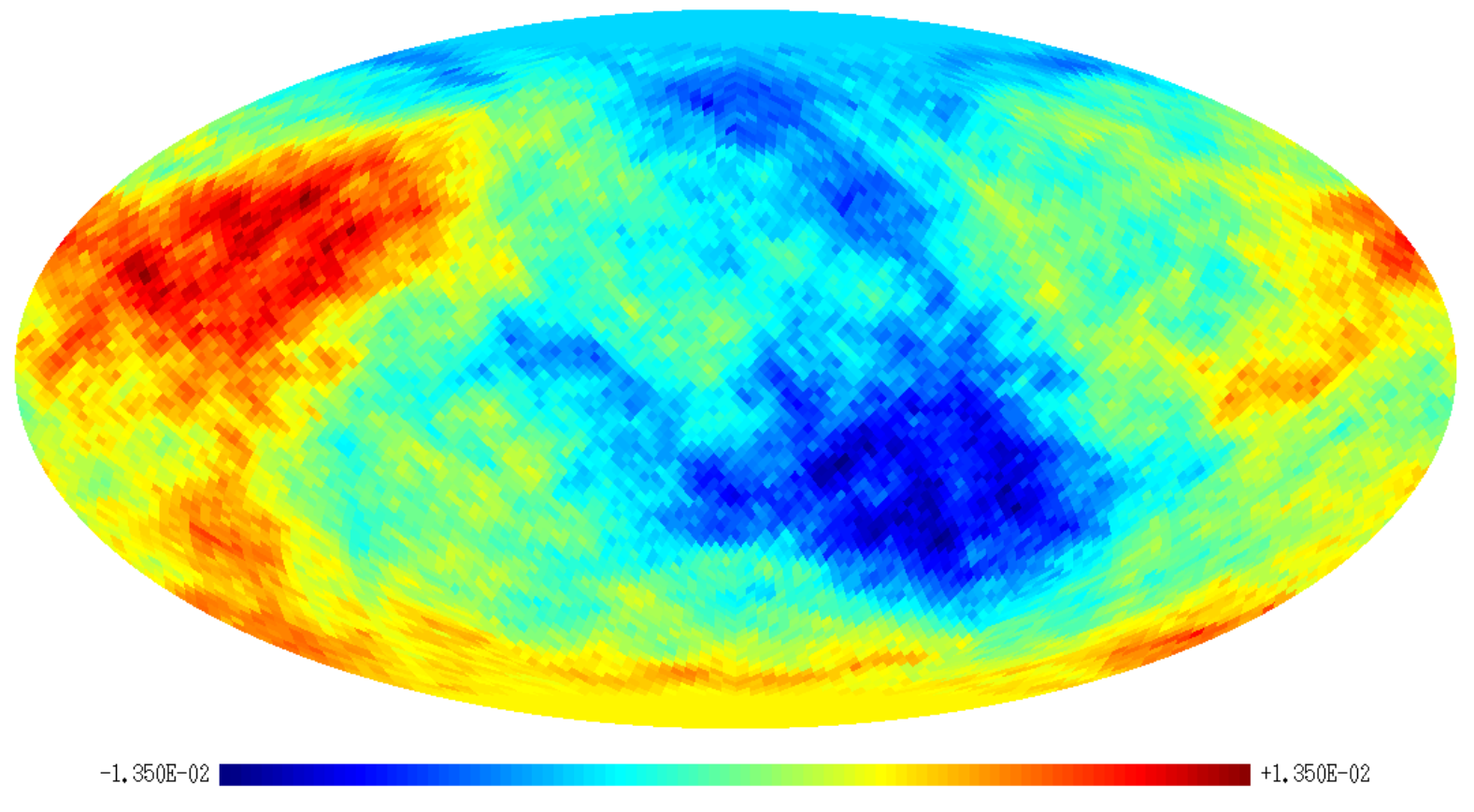}
     \caption{Map of $\Delta x$. Each pixel is coloured as per the value of $x_\text{up}-x_\text{down}$ when the z axis points through the centre of the pixel. This gives an indication of how the quantity varies across the sky. Catalogue C was used for this map.}
     \label{Fig:Delxmap}
 \end{figure}

The ratio of the sky brightness is also shown in Figure \ref{Fig:systematics}. Without any cuts or masking the NVSS data show very large spikes of high sky brightness. This is largely attributable to the bright and clustered sources near the galactic plane as is clearly visible in Figure \ref{Fig:NVSS}. The masking of the galactic plane with the flux cuts makes the variation of the sky brightness ratio fairly close to unity. For all our analysis in this work we have used the lower flux density cut of 20 mJy and upper flux density cut of 1000 mJy. Following examples in literature we understand that these measures would be sufficient to suppress the effects of systematics arising from the configuration differences of the survey. 

\section{Data Analysis}
\label{sec:analysis}
We find the best fit value of the spectral index $x$, by maximising the log-likelihood defined in equation (\ref{eq:likelihood}). This form does not take into account the error in flux density measurements and it is computationally cheap as opposed to the form in relation find (\ref{eq:L_w_error}). For computing equation  (\ref{eq:likelihood}), we need to determine $N_r$, $S_\text{max}$, $S_\text{min}$ and 
$\sum \ln S_i$. Then 
the calculation of the log-likelihood can be performed
which allows us to determine the spectral index $x$ in any selected region
of the sky.  
Since we need to perform the sum $\sum \ln S_i$ only once, the maximisation
of log-likelihood can be done with good computation speed. We simply scan
over all values of $x$ in a chosen range with a  
 very small step size, equal to $0.0001$. 
However for the form equation (\ref{eq:L_w_error}) we need to compute numerical integrals  for each source 
for every step in the change of spectral index. Given the very large number of sources this process is computationally prohibitive for our requirements. 
Hence we use this more extensive procedure only for a full sky calculation
of the spectral index. This allows us to  
to test the validity of our procedure based on the likelihood function
defined in equation (\ref{eq:likelihood}). We later discuss the differences in full sky fit of the spectral index obtained with the two procedures.

In our analysis, we are not considering the positional uncertainties in the NVSS catalogue. These uncertainties are not small but for our analysis the absolute position of the source is not relevant. In all our fits to the spectral index 
we work with a rather large area containing large number of sources. Hence if 
the source positions have errors in measurement then it would affect sources close to the edges of the area and the net effects should be small for our procedure. Hence we neglect the 
effects of positional errors for the present analysis.

\begin{figure}[t]
    \centering
    \includegraphics[width=1.0\textwidth]{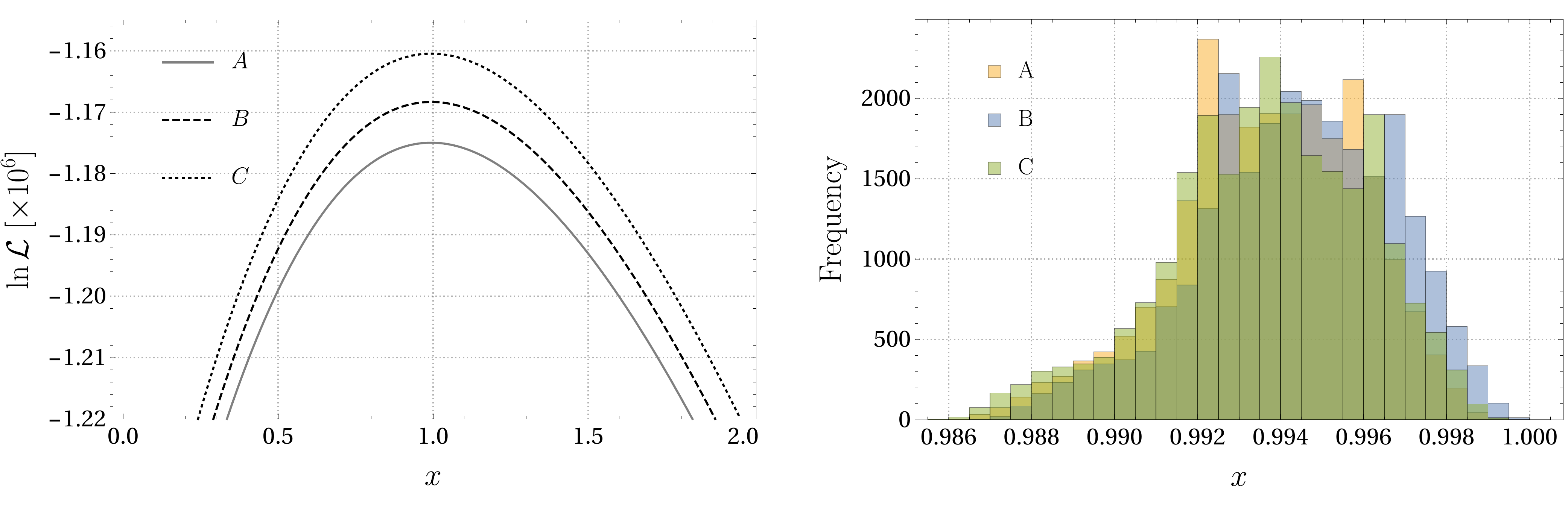}
    \caption{Left: The plot of variation of the log-likelihood with spectral index $x$ for the three different catalogues A, B and C. Right: The histogram for the half sky best fit values of the spectral index for the different catalogues.}
    \label{Fig:likehist}
\end{figure}

\subsection{Simulated data and method validation}
The method of finding the best fit value of spectral index is as described above. Since we have masked the data and imposed flux cuts we would like to test our likelihood maximisation technique against partial sky data and flux cuts. To ensure controlled testing of our analysis method we generate simulated data. 

To generate a simulated source catalogue, we used a Mathematica script. We use the probability distribution given by equation (\ref{eq:probability}) with a value of spectral index set from full sky fits, $S_\text{max}$ being a randomly chosen upper flux limit from the interval [1200, 1500] mJy and $S_\text{min}$ the randomly chosen lower flux limit from the interval [5, 15] mJy. 
A random number of source positions are generated which are between 1.9 million and 2.1 million and uniformly distributed over the full sky. The random positions are assigned a random flux density drawn from the defined probability distribution. 
This generates a catalogue of sources 
uniformly distributed over the sky and
obeying the distribution in equation (\ref{eq:distribution}). 

The catalogue simulated as above is in celestial coordinates. We then proceed to remove all sources south of declination $-40^\circ$. This produces a catalogue with 82\% sky coverage as the NVSS. We further impose the galactic plane masking and impose flux cuts of 20 mJy and 1000 mJy as discussed previously. We do not mask the 22 sites because such contaminations are not present and the area of these maskings are much smaller. We have followed the most important steps of our data processing pipeline to produce a masked partial sky data within a limited flux range. However, we do accept that this is an idealised simulation as we do not take into account the instrumental limitations of the VLA and the configurations used during the survey. Including such details would require us to incorporate much more information regarding the survey making the simulation computationally prohibitive for the purpose of this work. Testing against these simulations would help us find artefacts that might arise out of the data analysis pipeline. We essentially test the effects of flux density bounds and partial sky data.

We simulated a catalogue of 1971652 objects uniformly distributed over the sky with  a minimum flux of 9.87 mJy and maximum of 1471.96 mJy from a flux density distribution with an arbitrarily chosen value of the spectral index of 0.5. 
For the full sky simulated catalogue and no flux cuts we find the best fit spectral index to be 0.5009. When we mask 44\% of the sky to get identical sky coverage as the actual data sets prepared but without any flux cuts, we get spectral index of 0.5015. When analysing the data without any masking but 20-1000 mJy flux density bounds we obtain $x=0.4994$. Working with both the masking and flux density bounds leads to $x=0.4999$. 
So the deviation we find in the best fit value is 
very small, less than the error in the spectral index which is found to
about $0.2\%$. So our method of fitting is very robust to masking of the sky and the flux density bounds. We do not expect any bias in our results due to our data analysis pipeline.

\begin{table}[t]
\centering
\caption{The results for full sky best fit value of x, average of the best 
fit value x in each hemisphere, the maximum difference in the value of x in the upper and lower hemisphere and the direction of the upper hemisphere
corresponding to the maximum difference. As explained in text,
catalogue A only has galactic mask and flux bounds, B has additional 22 sites removed and C has another additional local sources removed. All the errors quoted are 1$\sigma$ errors obtained from the log-likelihood fit. For the hemisphere average values the quoted errors are the standard deviations of the hemisphere best fits.}
    \begin{tabular}{l c c c c c}
        \hline
        \hline
        \noalign{\vskip 0.05cm} 
        Data & Full sky $x$ &  Hemisphere Average $x$ & $\text{Maximum }\Delta x$ & (l,b)&p-value\\
        \hline 
        A & $0.9939 \pm 0.0025$ & $0.9938 \pm 0.0022$ & $0.0134 \pm 0.0050$ &$(122^\circ, 36^\circ )$& 2.76\%  \\
        B & $0.9946 \pm 0.0025$ & $0.9944 \pm 0.0023$ & $0.0131 \pm 0.0050$ &$(122^\circ, 36^\circ)$ & 3.69\%\\
        C & $0.9939 \pm 0.0025$ & $0.9937 \pm 0.0024$ & $0.0135 \pm 0.0050$ &$(122^\circ, 36^\circ)$& 2.76\% \\
        \hline 
    \end{tabular}
 
    \label{Table:results}
\end{table}

\subsection{Isotropy testing procedure}
We divide the sky into a HealPix \citep{Gorski:2005} grid of $N_\text{side}=32$. We vary the z axis of our system along the centre of each of the pixels. With z axis aligned along this direction we divide the sky in two hemispheres, one centred about $\hat z$ and the other along $-\hat z$. Then we find the value of the spectral indices $x_\text{up}$ and $x_\text{down}$ for the upper and lower hemispheres for z axis centred about each of the pixels of the HealPix grid. We define the quantity $\Delta x = x_\text{up} - x_\text{down}$. This gives us the difference in values of the spectral index in the two halves of sky. We then find the pixel (and its corresponding direction) for which $\Delta x$ is maximum. We have also checked if the full sky value of the spectral index matches to the average of all the half sky spectral indices. This is meant for checking consistency.

The number of sources and hence the error in the best fit value of 
 $x$ depend on the choice of hemisphere. 
In order to take this into account  
we define an error normalised quantity $\Delta_N$ as:
\begin{equation}
	\Delta_N = \frac{x_\text{up}-x_\text{down}}{\sqrt{\delta x_\text{up}^2+\delta x_\text{down}^2}},
\end{equation}
where $\delta x_\text{up}$ is the error in $x_\text{up}$ and $\delta x_\text{down}$, the error in $x_\text{down}$. The denominator is the two errors added in quadrature. Following the above process of finding the direction of maximum $\Delta x$, we find the direction along which $\Delta_N$ maximises.  

We also perform a fit to a dipole model for departure from isotropy. We write the spectral index $x$ for a pixel along a direction $\hat n$ as:
\begin{equation}
    x(\hat n) = x_0 + x_1 \hat \lambda \cdot \hat n,
    \label{eq:anisotropy}
\end{equation}
 where $\hat \lambda$ is the direction of maximum anisotropy. We fit for the four parameters of this model: one each for $x_0$ and $x_1$, and two for $\hat \lambda$. The fits are performed on catalogue C. 
As stated above, in this case our observable is the value of $x(\hat n)$ in
the pixel along the direction $\hat n$ using the HealPix pixelization
scheme. In this case we use  
the HealPix map with $N_\text{side}=2$ in order have sufficiently large number
of sources in any particular pixel. We avoided pixels with less than 4000 
sources. We finally ended up with 32 pixels, out of a total of 48,
with independent fits of the 
spectral index. The data set prepared in this way is then fitted to equation (\ref{eq:anisotropy}) by minimising $\chi^2$. The goodness of fit is
determined by comparison with a null case of fitting to an isotropic full 
sky spectral index.

\section{Results}
\label{sec:results}
The results corresponding to maximum difference $\Delta x$ are contained in Table \ref{Table:results}. The full sky value of the spectral index was fitted over the entire sky for each of the catalogues A, B and C. We searched for the pixel, for which the upper and lower hemispheres show maximum differences in the values of spectral indices, $\Delta x$. These maximum differences and the
 corresponding directions are given in Table \ref{Table:results}. 
The means of the spectral indices fitted over all the hemispheres are also given. The plots for the full sky log-likelihood are shown for the three catalogues in first panel of Figure \ref{Fig:likehist}. The histogram plots for the best fit spectral indices over the hemispheres over the entire HealPix grid are shown in the second panel of Figure \ref{Fig:likehist}. 
For consistency these values should be very close to the full sky average. 
The difference between the hemisphere average values and the corresponding 
full sky values is found to be very small. The quoted error in the values of the full sky fit of $x$ is $1\sigma$ error from the log-likelihood. The uncertainty in best fit values of $x$ is about 0.25\%. The standard deviation of the half sky best fit values of $x$ with the z axis along all possible positions of the HealPix grid is quoted as the error is hemisphere average value of $x$. 

We also see that the maximum $\Delta x$ is barely a 1.3\% variation from the full sky value. Hence, at best, the deviation in spectral index is a tiny effect. 
The error quoted in the value of $\Delta x$ is the maximum uncertainties in the values of $x_\text{up}$ and $x_\text{down}$, added in quadratures, 
for the direction along which $\Delta x$ is maximum. The uncertainty in $\Delta x$ is apparently very large, $\sim 37\%$. 
The variation of $\Delta x$ with direction is shown in  
Figure \ref{Fig:Delxmap}. The fit for the case of error normalised 
difference $\Delta_N$ also maximises along the same direction
 $(l=122^\circ, b=36^\circ)$ as that of the maximum $\Delta x$,
with a maximum $\Delta_N=2.69$ for catalogue C.

\begin{figure}[t]
    \centering
    \includegraphics[width=1.0\textwidth]{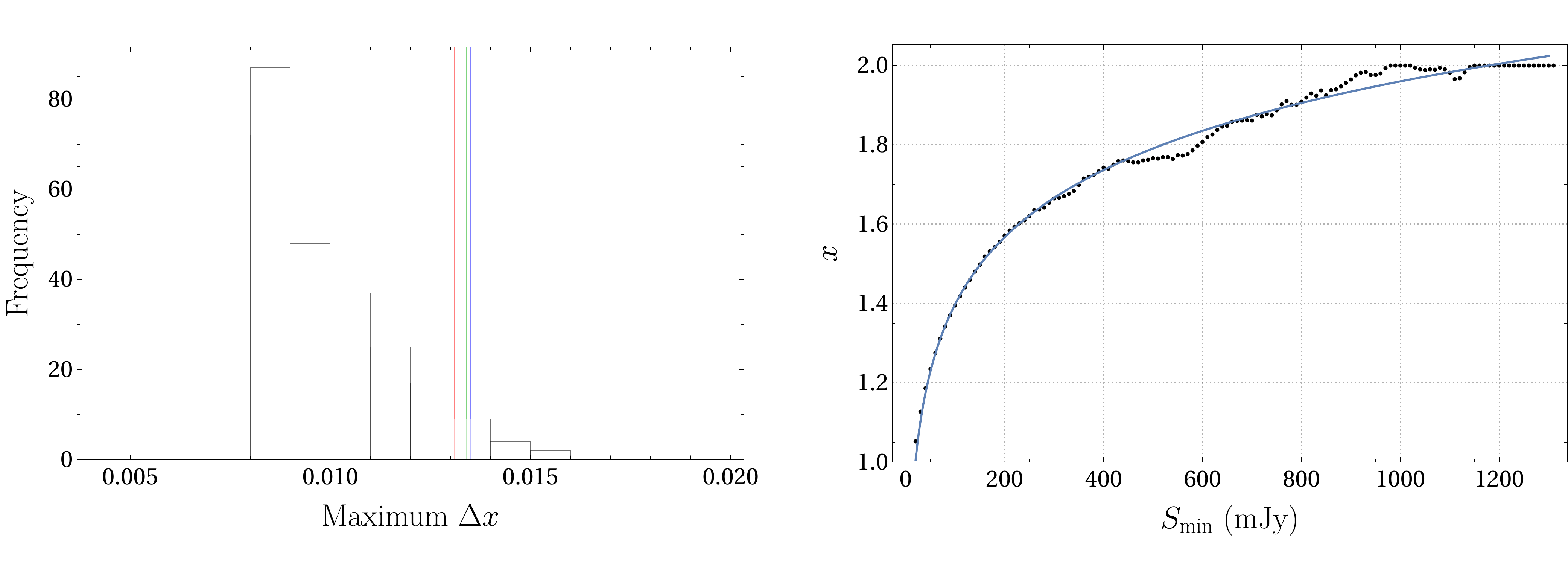}
    \caption{Left: Histogram plot of the results for maximum value of $\Delta x$ with 450 simulations with spectral index set at 0.9939. The green line indicates the result for catalogue A, red for catalogue B and blue for catalogue C. Right: This shows the variation in the spectral index with the lower flux cut $(S_\text{min})$. To study this variation we do not put any upper bound on the data of catalogue C and vary the lower flux density cut. The spectral index shows a logarithmic variation.} 
    \label{Fig:pvalue}
\end{figure}

To test the results we have simulated 450 catalogues as described previously with the spectral index of simulation set to 0.9939. As explained before all these simulated maps are processed by the entire data processing pipeline. 
The histogram plot for the values of maximum $\Delta x$ from these simulations 
is shown in the left graph of Figure \ref{Fig:pvalue}. We have also shown the observed values of maximum $\Delta x$ for all the catalogues in the figure. 
The figure shows that all these value lie towards the tail of the
histogram and have a relatively small probability. We calculate the p-value as the percentage of simulations for which the value of $\Delta x$ exceeds or equals the observed maximum value. The p-values for each of the results are also shown in Table \ref{Table:results}. Hence we find that the deviation from
isotropy is significant at 2 $\sigma$ level. This indicates a weak signal
of anisotropy which may be probed in future observations. 
Due to the relatively weak signal we conclude that the current data
are consistent with isotropy. 
The value of  $|\Delta x|$ is $0.0025 \pm 0.0049$ along the direction of the NVSS dipole \citep{Kothari:2013}. Along the direction of the CMB dipole 
modulation \citep{Eriksen:2004} $|\Delta x|$ is $0.0023 \pm 0.0050$.

We next describe our results for the dipole fit based on equation
\ref{eq:anisotropy}. In this case we determined the spectral index in each
pixel using $N_\text{side} = 2$. Each pixel was found to contain 
anywhere between 5000 and 9000 sources.  
 When the data were fitted with the isotropic null case  we get $x_0 = 0.9929 \pm 0.0036$ with a $\chi^2 = 27.5$. When fitted to our model of dipole anisotropy it gives: $x_0 = 0.9910 \pm 0.0039$, $x_1 = 0.0105 \pm 0.0081$ along ($l = 137^\circ \pm 37^\circ, b = 17 \pm 29^\circ$) with $\chi^2 = 24.1$.
Hence we do not find a significant signal of dipole anisotropy in the data.

We have thus far used the log-likelihood function defined in equation (\ref{eq:likelihood}), which does not incorporate the uncertainties in the measurement of the flux density. We have discussed before that the likelihood function defined in equation (\ref{eq:L_w_error}) is more complete because it incorporates the error in flux density measurements. We have computed a value for full sky best fit spectral index by using equation (\ref{eq:L_w_error}). The full sky best fit spectral index is found to be $0.9984 \pm 0.0022$ when including the error in flux density in the analysis. This is 0.45\% different from the full sky best fit values computed using equation (\ref{eq:likelihood}). This difference is not significant enough to require the use of computation intensive method of maximising the likelihood function defined in equation (\ref{eq:L_w_error}) for the purpose of this work. 

In our paper we have used a power law relationship given in equation 
\ref{eq:cumulative}. Although this is the form most widely used in the 
literature it is known that a generalization, i.e.,   
\begin{equation}
N_{>S}\propto S^{-x(S)}.
\end{equation}
with $x(S)=a+b\log S$ provides a much better fit \citep{Condon:1998,Kothari:2013,Tiwari:2013}. This is also clear from  
the second plot of Figure \ref{Fig:pvalue} where we have shown the spectral index plotted as a function of the lower flux density bound. We relaxed the upper flux density bound for obtaining this plot. It clearly shows a logarithmic
dependence of $x$ on $S$.  
We obtained a fit for $x(S)=a+b\log S$ with $a=0.2759 \pm 0.0124$ and $b=0.2438 \pm 0.0019$. It will clearly be of interest to extend our calculation 
in order to determine the direction dependence of both the parameters
$a$ and $b$ rather than restricting ourselves to a single parameter, i.e., the
spectral index
$x$. We postpone this to future research. 

\section{Conclusions}
In this work we have thoroughly tested the angular dependence of the 
spectral index $x$ with various flux density cuts and masking. We have searched for deviation of $x$ by dividing the sky in different hemispheres. 
Our technique of estimating the spectral index of a power law distribution was found to be robust with very small uncertainty in the fitted value. 
We do not 
find a significant deviation from isotropy. The deviation is at most a 
2$\sigma$ effect which is mildly interesting and should be probed in 
future observations. We conclude that 
if there is any departure in the isotropic behaviour of $\log N - \log S$ relation for the distant radio sources in the NVSS catalogue, it is indeed a tiny effect. 

\bibliographystyle{aasjournal}
\bibliography{bibliography}
\end{document}